\pdfoutput=1
\documentclass[11pt,a4paper]{article}
\usepackage[pdftex]{hyperref}
\usepackage{cleveref}
\usepackage{latexsym}
\usepackage{graphics}
\usepackage{amssymb}
\usepackage{graphicx}
\usepackage{float}
\usepackage{pifont}
\usepackage{cite}
\usepackage{pdfpages}

\usepackage{graphicx}
\usepackage{times}
\usepackage{tikz}
\usepackage{mathptm}
\usepackage{amsmath}
\usepackage{verbatim}
\usepackage{lineno,xcolor}
\usepackage{authblk}

\hypersetup{
    colorlinks = true,
    allcolors = {blue}
}

\providecommand{\pT}{$p_{\rm T}$ }

\providecommand{\raa}{$R_{\rm{\rm AA}}$ }

\providecommand{\vtwo}{$v_{\rm 2}$ }
\providecommand{\vn}{$v_{n}$ }

\providecommand{\s}{$\sqrt{s}$ }
\providecommand{\snn}{$\sqrt{s_{\rm NN}}$ }

\if@twoside   
\else         
\fi

\begin{document}
\title{Elliptic flow of electrons from beauty-hadron decays extracted from Pb--Pb collision data at \snn = 2.76 TeV}

\author[1]{D. Moreira de Godoy}
\author[1]{F. Herrmann}
\author[1]{M. Klasen}
\author[1,2]{C. Klein-B\"osing}
\author[3]{A. A. P. Suaide}

\affil[1]{Westf\"alische Wilhelms-Universit\"at M\"unster}
\affil[2]{ExtreMe Matter Institute EMMI,  GSI Helmholtzzentrum f\"ur Schwerionenforschung}
\affil[3]{Universidade de S\~ao Paulo}


\date{}

\maketitle

\begin{abstract}
We present a calculation of the elliptic flow of electrons from beauty-hadron decays in semi-central Pb--Pb collisions at centre-of-mass energy per colliding nucleon pair, represented as  $\sqrt{s_{\rm NN}}$,  of 2.76 TeV. The result is obtained by the subtraction of the charm quark contribution in the  elliptic flow of electrons from heavy-flavour hadron decays in  semi-central Pb--Pb collisions at \snn = 2.76 TeV recently made publicly available by the ALICE collaboration.
\end{abstract}

\section{Introduction}
\label{Sec:Introduction}
 
The nuclear matter exposed to conditions of high temperature and energy density is expected to undergo a phase transition to a colour deconfined state of matter~\cite{Karsch:2001cy,pQCD5}, the Quark-Gluon Plasma (QGP). The conditions for the phase transition can be achieved in the laboratory with collisions of heavy ions at high energies~\cite{Karsch:2001cy}. 
The properties of the medium formed in the laboratory can be probed with a unique degree of control by particles from decays of heavy flavours (charm and beauty), since heavy quarks are mainly produced in hard parton scattering processes~\cite{Norrbin2000,Cacciari2012,Klasen:2014dba}  at the initial stage of heavy-ion collisions~\cite{PhysRevC.77.024901,0954-3899-34-8-S36} and participate in the entire evolution of the created system. The partons traversing the medium lose energy via collisional and radiative processes~\cite{Collc,Djordjevic:2006tw,Wicks:2007am,Baier:1996kr,PhysRevC.77.024905} in the interaction with the medium constituents.
The energy loss of partons  is predicted to be dependent on their colour charge and mass, resulting in a hierarchy where beauty quarks lose less energy than charm quarks, charm quarks lose less energy than light quarks, and quarks lose less energy than gluons~\cite{Wicks:2007am,Dokshitzer:2001zm,PhysRevD.69.114003}. 
The heavy-flavour energy loss can be investigated experimentally with the nuclear modification factor   ($R_{\rm AA}$) of heavy-flavour particles, defined as 

\begin{equation}
R_{\rm{AA}} = \frac{1}{\left\langle T_{\rm{AA}} \right\rangle} \frac{\mathrm{d}N_{\rm{AA}}/\mathrm{d}p_{\rm T}}{\mathrm{d}\sigma_{\rm{pp}}/\mathrm{d}p_{\rm T}}, 
\label{eq:raa}
\end{equation}

\noindent where $\mathrm{d}N_{\rm{AA}}/\mathrm{d}p_{\rm T}$ is the transverse momentum ($p_{\rm T}$) differential yield in nucleus-nucleus (AA) collisions; $ \langle T_{\rm{AA}} \rangle$ is the average nuclear overlap function in nucleus-nucleus collisions, given by the ratio of the average number of binary collisions and the inelastic cross section; and  
$\mathrm{d}\sigma_{\mathrm{pp}}/\mathrm{d}p_{\rm T}$  is the $p_{\rm T}$-differential cross section in proton-proton (pp) collisions. 
A suppression of the yield of $\mathrm{D}$ mesons and leptons from heavy-flavour hadron decays    ($R_{\rm AA}<$ 1) for $p_{\rm T} > $ 3 GeV/$c$ was observed in gold-gold (Au--Au) collisions at \snn = 200 GeV at the Relativistic Heavy Ion Collider (RHIC)~\cite{STARRaaeAuAu,STARRaaD0AuAu,PHENIXRaaeAuAu,PHENIXRaaeAuAubis} and in lead-lead (Pb--Pb) collisions at \snn = 2.76 and 5.02 TeV at the Large Hadron Collider (LHC)~\cite{ALICEDRPbPb,ALICEDRPbPbcent,ALICEmuRPbPb,CMSD0Raa,Adam:2016wyz,HFERAA,ATLAS-CONF-2015-053}, indicating energy loss of heavy flavours in the medium. 
An experimental hint to the quark mass dependence of the heavy-flavour energy loss has been found in the comparison of the \raa of $\mathrm{D}$ mesons and non-prompt J/$\psi$ from $\mathrm{B}$-hadron decays in central Pb--Pb collisions at \snn = 2.76 TeV at the LHC~\cite{ALICEDRPbPbcent,CMSnpjpsi,Khachatryan:2016ypw}.  The observed difference of these measurements is described by model calculations~\cite{Djordjevic2014298,Andronic:2015wma} as predominantly due to the quark mass dependence of the parton energy loss.

The interaction of heavy quarks with the medium can be further investigated with the  azimuthal anisotropy of  heavy-flavour particles, extracted from the coefficients \vn of the Fourier decomposition of the particle azimuthal distribution in the transverse plane~\cite{PhysRevC.58.1671}

\begin{equation}
\frac{\mathrm{d}N}{\mathrm{d}\left( \varphi - \Psi_{n} \right) } \propto  1+ 2 \sum_{n=1}^{\infty}  v_{n}\cos\left[n\left( \varphi - \Psi_{n} \right)\right],
\label{eq:FourierSeries}
\end{equation}   

\noindent  where $\varphi$ is the azimuthal angle of the heavy-flavour particles and $\Psi_{n}$ is the symmetry-plane angle of the $n^{th}$-order harmonic. 
The second Fourier coefficient $v_2$, called elliptic flow,  quantifies the elliptic azimuthal anisotropy of the emitted particles. 
The origin of the elliptic azimuthal anisotropy of heavy-flavour particles in non-central heavy-ion collisions depends on the transverse-momentum interval. While the \vtwo at low $p_{\rm T}$ is sensitive to the collective motion of the medium constituents caused by pressure gradients, the \vtwo at high $p_{\rm T}$ can constrain the path-length dependence of the in-medium energy loss of heavy quarks, resulting from the direction of the particles that traverse the ellipsoidal nuclear overlap region.
The elliptic flow of prompt D mesons at mid-rapidity is observed to be positive in  30--50\% Pb--Pb collisions at \snn = 2.76 TeV at the LHC~\cite{Abelev:2013lca,Abelev:2014ipa} with 5.7$\sigma$ significance in the interval 2 $< p_{\rm T} <$ 6 GeV/$c$, indicating that charm quarks participate in the collective motion of the system. Measurements of the prompt D-meson $v_2$ in Pb--Pb collisions at \snn = 5.02 TeV~\cite{CMS:2016jtu,Acharya:2017qps} have smaller uncertainties compared to the ones in Pb--Pb collisions at \snn = 2.76 TeV. The results at the two  collision energies are compatible within uncertainties. 
The prompt D$_ {s}^{+}$ \vtwo in semi-central  Pb--Pb collisions at \snn = 5.02 TeV is compatible within uncertainties with the average  of prompt D$^{0}$, D$^{+}$, and D$^{*+}$  \vtwo in the same collision system~\cite{Acharya:2017qps}.
A positive \vtwo is also observed for leptons from heavy-flavour hadron decays at low and intermediate \pT in semi-central Au--Au collisions at \snn = 200 GeV at RHIC~\cite{Adare:2010de,Adamczyk:2014yew} and in semi-central Pb--Pb collisions at \snn = 2.76 TeV at the LHC~\cite{Adam:2016ssk,Adam:2015pga,ATLAS-CONF-2015-053}.  In particular, the \vtwo of electrons from heavy-flavour hadron decays is observed to be positive with 5.9$\sigma$ significance in the range 2 $< p_{\rm T} <$ 2.5 GeV/$c$ in 20--40\%  Pb--Pb collisions at \snn = 2.76 TeV.

In view of the experimental results on the elliptic flow of heavy-flavour particles, an important  question that remains open is whether beauty quarks take part in the collective motion in the medium. 
The first measurement of the \vtwo of non-prompt J/$\psi$ mesons from $\mathrm{B}$-hadron decays is compatible with zero within uncertainties in two kinematic regions, 
6.5 $< p_{\rm T} <$ 30 GeV/$c$ and $|y| <$ 2.4, and   3 $< p_{\rm T} <$ 6.5 GeV/$c$ and 1.6 $< |y| <$ 2.4, in 10--60\% Pb--Pb collisions at \snn = 2.76 TeV at the LHC~\cite{Khachatryan:2016ypw}.
In this paper, we present a method to subtract the contribution of  charm quarks in the published measurement of the elliptic flow of electrons from heavy-flavour hadron decays in semi-central Pb--Pb collisions at \snn = 2.76 TeV performed by the ALICE collaboration. The calculation uses  as input the \vtwo coefficients of prompt D mesons and electrons from heavy-flavour hadron decays measured by the ALICE collaboration~\cite{Abelev:2014ipa,Adam:2016ssk} and three different results for the relative contribution of electrons from beauty-hadron decays to the yield of electrons from heavy-flavour hadron decays~\cite{Abelev:2014hla,2016507,Uphoff:2014hza,Uphoff:2012gb}.

\section{Methodology}
\label{Sec:Methodology}

The particle azimuthal distribution of electrons from heavy-flavour hadron decays ($e \leftarrow c+b$)   can be separated into the contributions of electrons from charm-hadron decays ($e \leftarrow c$) and from beauty-hadron decays ($e \leftarrow b$). Consequently, the elliptic flow of electrons from beauty-hadron decays can be expressed as

\begin{equation}
v_{2}^{e \leftarrow b} = \frac{v_{2}^{e \leftarrow c+b} - (1- R)v_{2}^{e \leftarrow c}}{R},
\label{eq:ElecFrombeautyV2}
\end{equation}   

\noindent where $R$ represents the relative contribution of electrons from beauty-hadron decays to the yield of electrons from heavy-flavour hadron decays.

In the following, we present the currently published measurements and, in case there is no available measurement, our calculations  of the three observables required to obtain the elliptic flow of electrons from beauty-hadron decays. 
Based on   available results on open heavy flavours  at RHIC and LHC, the most suitable system for this analysis is  the Pb--Pb collision system at \snn = 2.76 TeV in the 20--40\% centrality class, which corresponds to the centrality range where the measured \vtwo of electrons from heavy-flavour hadron decays is observed to be positive with a maximum significance~\cite{Adam:2016ssk} and thus a possible elliptic flow of electrons from beauty-hadron decays is expected to be more significant.
In this analysis, the \vtwo and \raa of heavy-flavour particles are assumed to be the same at slightly different mid-rapidity ranges ($|y| <$ 0.5, 0.7 and 0.8)  in which the measurements needed in the calculation are available. Indeed, no dependence on rapidity was observed in recent ALICE results on those observables for electrons from heavy-flavour hadron decays at mid-rapidity ($|y| <$ 0.7 for \vtwo and $|y| <$ 0.6 for \raa measurements) and muons from heavy-flavour hadron decays at forward   rapidity (2.5$ < y < $4)~\cite{Adam:2016ssk,HFERAA}.

\subsection{Elliptic flow of electrons from heavy-flavour hadron decays}
\label{Sec:ElectronFromHFV2}

The result on the elliptic flow of electrons from heavy-flavour hadron decays ($v_{2}^{e \leftarrow c+b}$) at  mid-rapidity  ($|y| < 0.7$) in 20--40\% Pb--Pb collisions at \snn = 2.76 TeV published by the ALICE collaboration~\cite{Adam:2016ssk}  is used in this analysis. The $v_{2}^{e \leftarrow c+b}$ is measured in the interval 0.5 $< p_{\rm T} <$ 13 GeV/$c$ with the event plane method~\cite{PhysRevC.58.1671}. A positive value is observed in the interval  2 $< p_{\rm T} <$ 2.5 GeV/$c$ with significance of 5.9$\sigma$~\cite{Adam:2016ssk}.

\subsection{Relative contribution of electrons from beauty-hadron decays to the yield of electrons from heavy-flavour hadron decays}
\label{Sec:R}

The measurement of the relative contribution of electrons from beauty-hadron decays to the yield of electrons from heavy-flavour hadron decays ($R$) has been published by the ALICE collaboration only in pp collisions at \s = 2.76 TeV~\cite{Abelev:2014hla,2016507}.
The coefficient $R$ measured in pp collisions is used in the analysis with the caveat that initial- and final-state effects modify the yield of electrons from heavy-flavour hadron decays in heavy-ion collisions. In particular, the coefficient $R$ at high $p_{\rm T}$ is expected to be higher in Pb--Pb collisions compared to pp collisions, since the in-medium energy loss of charm quarks is predicted to be larger than the one of beauty quarks~\cite{Adam:2016wyz}. 
Therefore, the factor $R$ at high $p_{\rm T}$  in Pb--Pb collisions at \snn = 2.76 TeV is expected to have an exclusive value between  the measured factor $R$ in pp collisions at \s = 2.76 TeV and unity. Consequently, according to Eq.~\ref{eq:ElecFrombeautyV2}, 
the minimum value of the  \vtwo of electrons from beauty-hadron decays can be computed with the $R$ measured in pp collisions.

\begin{figure}
  \centering
  \includegraphics[height=0.45\textwidth]{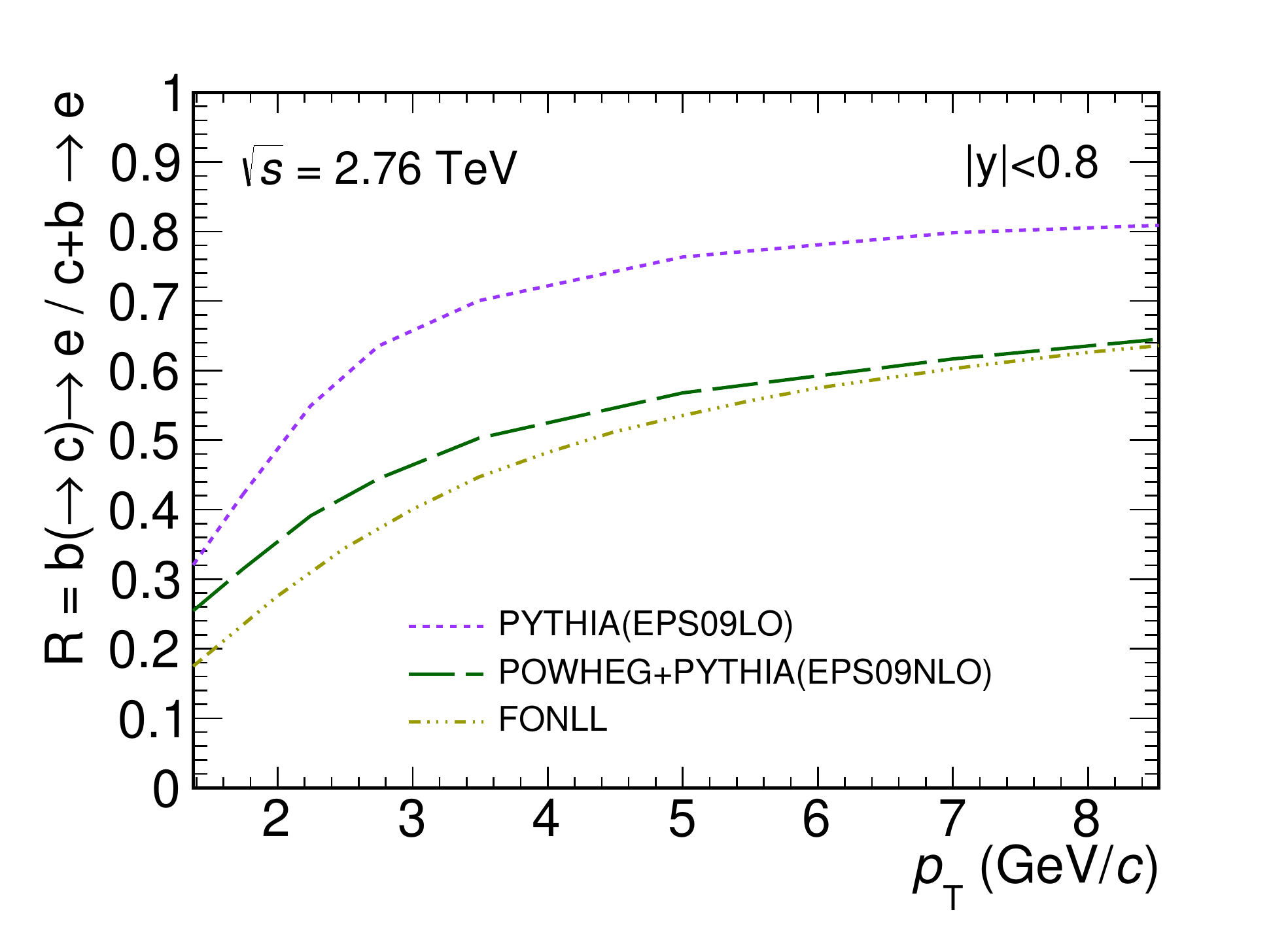}
   \caption{Comparison of the relative contribution of electrons from beauty-hadron decays to the yield of electrons from heavy-flavour hadron decays ($R$) at \s = 2.76 TeV obtained with POWHEG+PYTHIA~\cite{POWHEG,Sjostrand:2007gs,Pythia8} at NLO accuracy using  EPS09 NLO nPDFs, with PYTHIA at LO accuracy  using EPS09 LO nPDFs, and with FONLL calculation~\cite{1126-6708-1998-05-007,Cacciari2012} using CTEQ6.6 PDFs. Only the central values are shown.}
   \label{fig:POWHEG}
\end{figure}

In addition to the available measurement, the coefficient $R$ is obtained with a Monte Carlo (MC) simulation based on POWHEG~\cite{POWHEG}, which provides the calculation of the heavy-flavour production in hadronic collisions at Next-to-Leading Order (NLO) accuracy. The POWHEG results are interfaced to  PYTHIA~\cite{Sjostrand:2007gs,Pythia8}  in order to generate the shower, hadronisation and decay. 
In agreement with other heavy-flavour production tools, e.g. pQCD calculation at Fixed Order plus Next-to-Leading Logarithms (FONLL)~\cite{1126-6708-1998-05-007,Cacciari2012} and earlier pp calculations~\cite{Klasen:2014dba}, the square root of the quadratic sum of the quark mass ($m_Q$) and $p_{\rm T}$  are used as renormalization and factorization scales, i.e. $\mu=\mu_f=\mu_r=\sqrt{m_Q^2+p_{\rm T}^{2}}$. The charm- and beauty-quark masses are set as 1.5 GeV and 4.75 GeV, respectively.
Even though the calculated coefficient $R$ is sensitive to the choice of heavy-quark masses and scales, only the central value is shown in this analysis.
 Admittedly, the described framework is designed for pp collisions, but by making use of the EPS09~\cite{EPS09} NLO nuclear Parton Distribution Functions (nPDFs) the framework is able to account for initial-state cold nuclear effects.
The nPDF gluon shadowing results in reduced $p_{\rm T}$-differential cross sections for electrons from heavy-flavour hadron decays for  \pT $<$ 6 GeV/$c$ and affects contributions from charm quarks stronger than those from beauty quarks. 
Thus, it provides a lower baseline for the factor $R$, which is suggested to be further enhanced by medium interactions as it will be discussed in this paper. 
In addition, the $R$ coefficient is also obtained with a leading order (LO) calculation based on PYTHIA using EPS09 LO nPDFs to study the impact of NLO corrections~\cite{Klasen:2014dba}.
The comparison of the calculations with LO and NLO approaches, shown in  Fig.~\ref{fig:POWHEG}, reveals that 
the factor $R$ is reduced with the NLO corrections, stressing the importance of NLO calculations. In fact, the additional processes of heavy-flavour production at NLO  give rise to large logarithmic corrections to the charm- and beauty-quark cross sections depending on the heavy-flavour mass. 
The corresponding FONLL calculation of the factor $R$ using CTEQ6.6 PDFs, which is also shown in Fig.~\ref{fig:POWHEG}, is similar to the POWHEG+PYTHIA   (EPS09NLO) result at high $p_{\rm T}$.

The result on the factor $R$ from the BAMPS heavy-flavour transport model~\cite{Uphoff:2014hza,Uphoff:2012gb}, which includes  collisional and radiative in-medium energy loss of heavy quarks, is also employed in the analysis to obtain the \vtwo of electrons from beauty-hadron decays. The choice of the BAMPS model  is justified by the good agreement of the predictions for the \raa of electrons from beauty- and heavy-flavour hadron decays  for $p_{\rm T} > $ 3 GeV/$c$ in central  Pb--Pb collisions at \snn = 2.76 TeV with what measured by the ALICE collaboration~\cite{HFERAA,Adam:2016wyz}.

In this analysis, the  $R$ coefficient measured in pp collisions using the electron-hadron azimuthal technique~\cite{Abelev:2014hla,2016507} and the ones obtained with POWHEG+PYTHIA  (EPS09NLO) and with the BAMPS model are used to estimate the elliptic flow of electrons from beauty-hadron decays. The $R$ coefficient measured in pp collisions is also obtained with the track impact parameter method~\cite{Abelev:2014hla,2016507}. Results obtained with both techniques are compatible within uncertainties.

\subsection{Elliptic flow of electrons from charm-hadron decays}
\label{Sec:ElectronFromCharmV2}

The elliptic flow of electrons from charm-hadron decays ($v_{2}^{e \leftarrow c}$) is estimated using a MC simulation  of decays of D$^{0}$ mesons into electrons with PYTHIA. The MC simulation is based on two observables measured for D$^{0}$ mesons in Pb--Pb collisions at \snn = 2.76 TeV: 

\begin{itemize}
\item[--] the  $p_{\rm T}$-differential yield, which is used as a probability distribution  for finding a D$^{0}$ meson with a certain $p_{\rm T}$; 
\item[--] the $p_{\rm T}$-differential $v_{2}$, which is used to obtain the   $\varphi_{\rm{D}^{0}} - \Psi_{2}$ probability distribution with  Eq. \ref{eq:FourierSeries}.
\end{itemize}

\begin{figure}
\centering
  \includegraphics[height=0.355\textwidth]{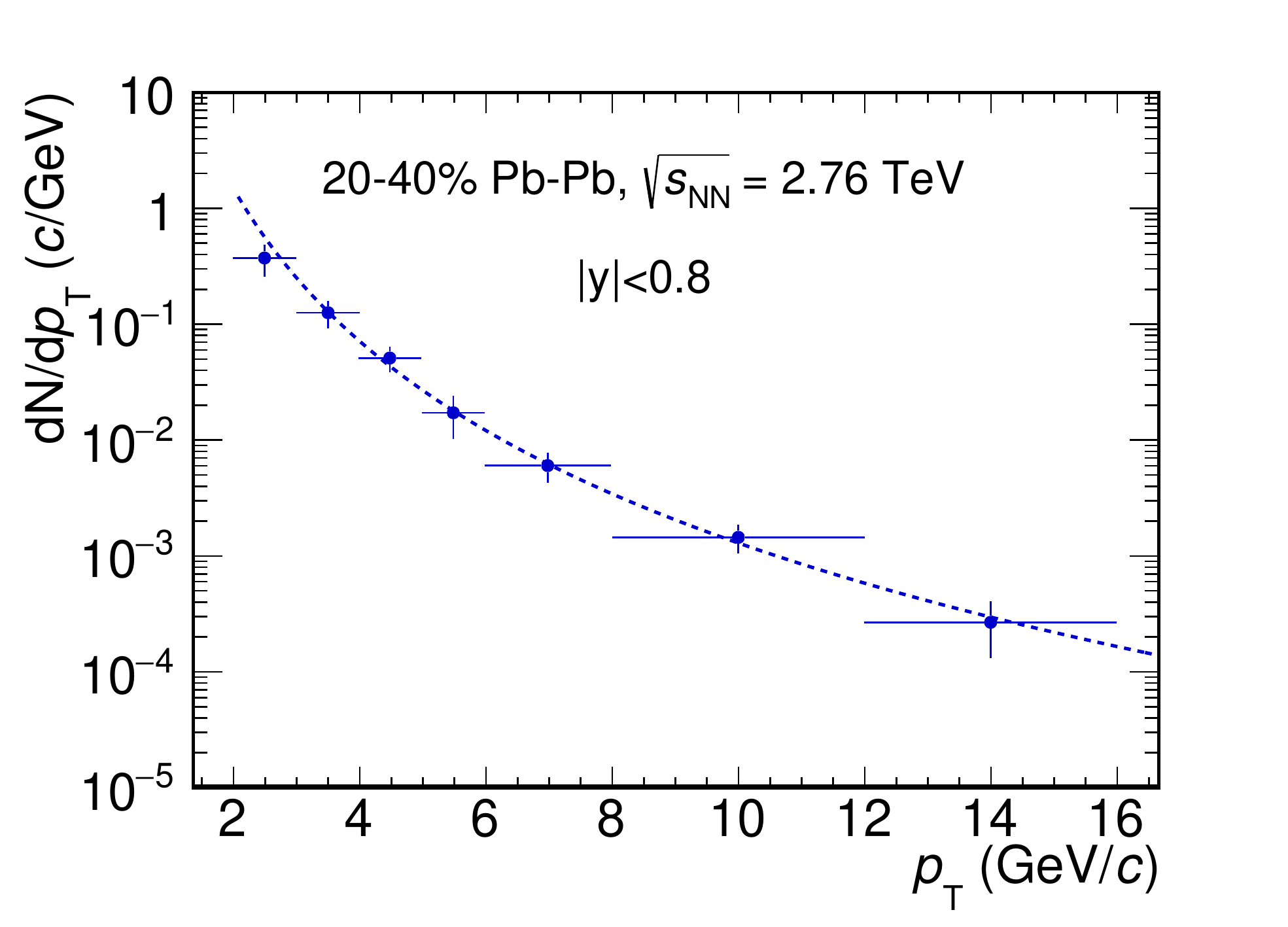}
    \includegraphics[height=0.355\textwidth]{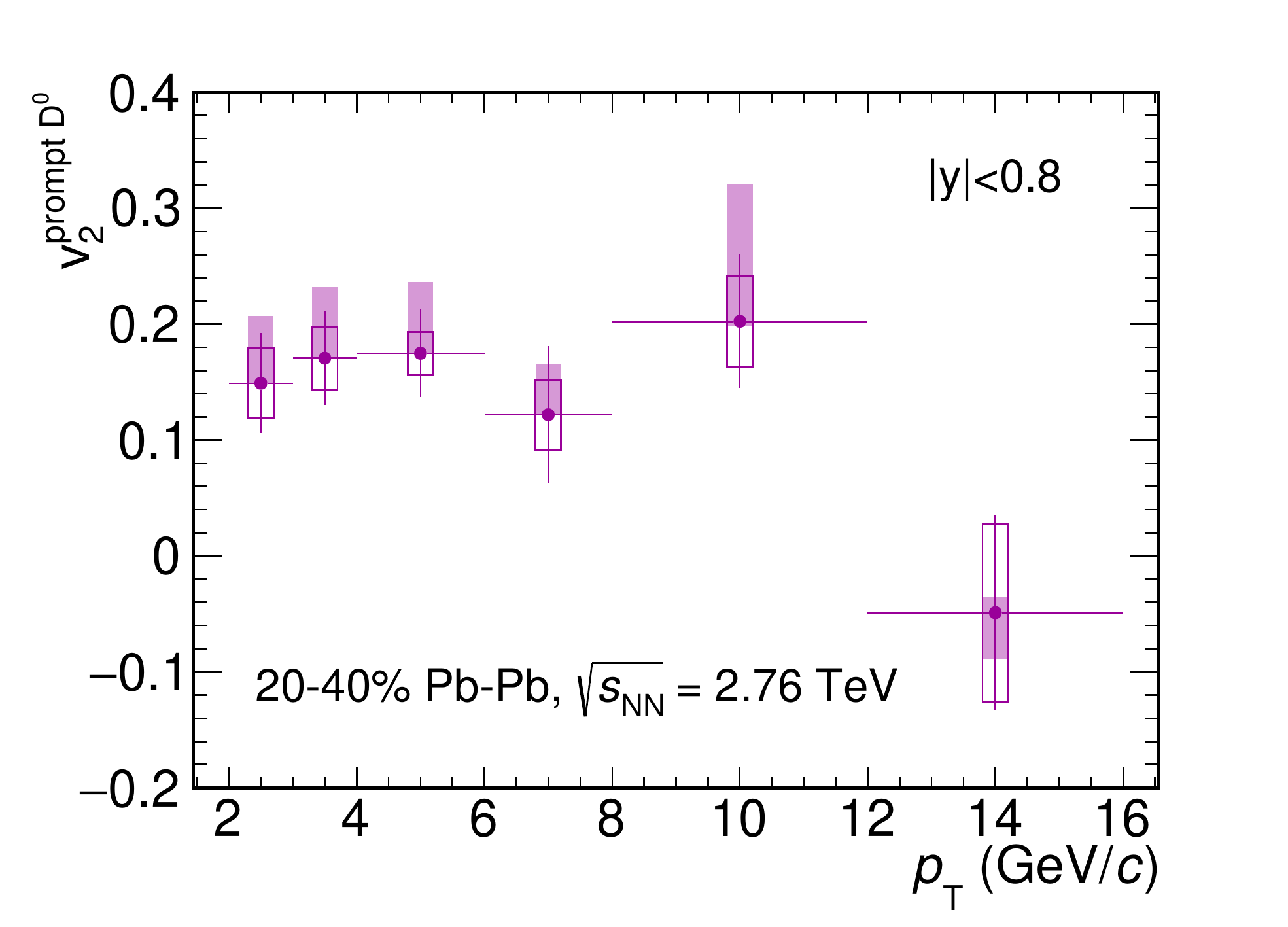}
  \caption{Left:  Estimated $p_{\rm T}$-differential yield of D$^{0}$ mesons  in 20--40\%   Pb--Pb collisions at \snn = 2.76 TeV from ALICE results~\cite{ALICEDRPbPb}. The dashed line corresponds to the power-law fitted function. Right:  Estimated \vtwo of prompt D$^{0}$ mesons  in 20--40\%   Pb--Pb collisions at \snn = 2.76 TeV from ALICE results~\cite{Abelev:2014ipa}. The vertical error bars represent the statistical uncertainties and the horizontal error bars indicate the bin widths. The empty and filled boxes represent the systematic uncertainties from data and from the B feed-down subtraction, respectively, in the D$^0$-meson \vtwo measurement~\cite{Abelev:2014ipa}.}
   \label{fig:CockInput}
\end{figure}

In fact, the $p_{\rm T}$-differential yield of D$^{0}$ mesons  at mid-rapidity ($|y| <$ 0.8) in 20--40\% Pb--Pb collisions at \snn = 2.76 TeV is estimated from the ALICE results on the $p_{\rm T}$-differential yield  and \raa  of  prompt D$^{0}$ mesons at mid-rapidity ($|y| <$ 0.5) in 0--20\% Pb--Pb collisions at the same collision energy~\cite{ALICEDRPbPb} as 

\begin{equation}
\small
\left( \frac{\mathrm{d}N_{\rm AA}}{\mathrm{d}p_{\rm T}} \right)^{20-40\%}  =  C_{\Delta y} \   C_{\langle T_{\rm AA} \rangle}   \ C_{R_{\rm AA}}  
 \ \left( \frac{\mathrm{d}N_{\rm AA}}{\mathrm{d}p_{\rm T}} \right)^{0-20\%},
\label{eq:pTSpectrumD}
\end{equation} 

\noindent where the coefficient $C_{\Delta y}$ = 1.6 corresponds to the scaling factor of the yield from $|y| <$ 0.5 to $|y| <$ 0.8 in pp collisions, assuming a uniform distribution of the D$^{0}$-meson yield within the rapidity range. The terms $C_{\langle T_{\rm AA} \rangle}$ = 0.362 $ \pm$ 0.020~\cite{ALICEDRPbPb} and $C_{R_{\rm AA}}$ are the ratios of the average nuclear overlap function and the D$^{0}$-meson $R_{\rm AA}$, respectively, in Pb--Pb collisions at \snn = 2.76 TeV in  the 20--40\% centrality class to the ones in the 0--20\%  centrality class.
Note that the  terms $C_{\Delta y}$ and $C_{\langle T_{\rm AA} \rangle}$   are constant, so they do not play a role in the determination of the D$^{0}$-meson \pT probability distribution.
The non-measured \raa of D$^{0}$ mesons in 20--40\% Pb--Pb collisions at \snn = 2.76 TeV is  estimated by the average of the ALICE results on the D$^{0}$-meson $R_{\rm AA}$ in Pb--Pb collisions at the same collision energy in the 0--20\% and 40--80\% centrality classes~\cite{ALICEDRPbPb} weighted by the  corresponding yield of D$^{0}$ mesons in each centrality class.
The  resulting \pT distribution of D$^{0}$ mesons in 20--40\% Pb--Pb  collisions obtained from Eq.~\ref{eq:pTSpectrumD} is then fitted by a power-law function (left panel of Fig.~\ref{fig:CockInput}), considering  the statistical uncertainty of the experimental results. The fit function is used as the D$^{0}$-meson \pT probability distribution.

The \vtwo of prompt D$^{0}$ mesons in 20--40\% Pb--Pb collisions at \snn = 2.76 TeV (right panel of Fig.~\ref{fig:CockInput}) is obtained by the  arithmetic average  of the measured prompt D$^0$-meson  \vtwo in Pb--Pb collisions at the same collision energy in the 10--30\% and 30--50\% centrality classes~\cite{Abelev:2014ipa}. Indeed, experimental results show that the \vtwo of heavy-flavour particles increases with the centrality class~\cite{Abelev:2014ipa, Adare:2010de, Adam:2016ssk,Adam:2015pga}, which is consistent with the qualitative expectation of increasing of the elliptic anisotropy from central to peripheral nucleus-nucleus collisions.
The statistical and systematic uncertainties are propagated considering the prompt D$^0$-meson  \vtwo  in the 10--30\% and 30--50\% centrality classes as uncorrelated as a conservative estimation.
In the D$^0$-meson \vtwo measurement by the ALICE collaboration, the central value was obtained by assuming that  the \vtwo coefficients of prompt D mesons and D mesons from B-meson decays are the same~\cite{Abelev:2014ipa}. However, the systematic uncertainty related to this assumption, referred to as systematic uncertainty from the B feed-down subtraction, was evaluated by the ALICE collaboration. It was assumed that the 
\vtwo of prompt D mesons from B-meson decays should be between zero and \vtwo of prompt D mesons, resulting in the upper and lower limits of the systematic uncertainty, respectively. Therefore, the B feed-down contribution decreases the absolute value of the D$^0$-meson \vtwo and thus the systematic uncertainty is restricted to the upper (lower) limit when the \vtwo is positive (negative). Since the measured D$^0$-meson \vtwo coefficients are negative in the 8 $< p_{\rm T} <$ 12 GeV/$c$ and 12 $< p_{\rm T} <$ 16 GeV/$c$ intervals in the 10--30\% and 30--50\% centrality classes, respectively, the resulting propagated systematic uncertainty from the B feed-down subtraction contains lower and upper limits.

Finally, the estimated \pT and \vtwo distributions of D$^{0}$ mesons  in 20--40\%   Pb--Pb collisions at \snn = 2.76 TeV are used to obtain the $v_{2}^{e \leftarrow c} = \langle \cos \left[ 2 \left( \varphi_{e} - \Psi_{2} \right) \right] \rangle$ in the same collision system using the PYTHIA event generator.
The azimuthal angle of electrons  ($\varphi_{e}$) takes into account the angular separation between electrons and their parent D$^{0}$ mesons.

\subsubsection{Statistical uncertainty}

The statistical uncertainty of the  D$^{0}$-meson $v_{\rm 2}$ is used as input for the MC simulation  to obtain the statistical uncertainty of the  $v_{2}^{e \leftarrow c}$. The statistical uncertainties of the measurements used to obtain the $p_{\rm T}$-differential yield of D$^{0}$ mesons are considered in the fit of the D$^{0}$-meson probability distribution. Further variations are considered as systematic uncertainties.


\subsubsection{Systematic uncertainty}

The systematic uncertainties from data and from the B feed-down subtraction of the D$^{0}$-meson $v_{\rm 2}$ (right panel of Fig.~\ref{fig:CockInput}) are  used as input for the MC simulation to obtain the systematic uncertainty of the  $v_{2}^{e \leftarrow c}$. The following is a discussion on other sources of systematic uncertainty that can influence the $v_{2}^{e \leftarrow c}$ estimation.

In order to validate the Eq. \ref{eq:pTSpectrumD}, the $p_{\rm T}$-differential yield  and \raa  of  prompt D$^{0}$ mesons in 40--80\% Pb--Pb collisions at \snn = 2.76 TeV~\cite{ALICEDRPbPb} are also used as reference to obtain the  $p_{\rm T}$-differential yield of D$^{0}$ mesons in the 20--40\% centrality class. The result is the same as the one obtained with the 0--20\% centrality class (left panel of Fig.~\ref{fig:CockInput}).

The ALICE result on the D$^{0}$-meson $R_{\rm AA}$ in the 30--50\% centrality class~\cite{Abelev:2014ipa} is  used as an alternative for the D$^{0}$-meson \raa estimation in the 20--40\% centrality class. No significant difference is observed in the resulting $v_{2}^{e \leftarrow c}$ with respect to the one obtained with the $R_{\rm AA}$  estimated by the average of the D$^{0}$-meson $R_{\rm AA}$ measurements in the 0--20\% and 40--80\% centrality classes weighted by the  corresponding yield of D$^{0}$ mesons in each centrality class.

The systematic uncertainties of the measurements of the $p_{\rm T}$-differential yield  and \raa  of  prompt D$^{0}$ mesons are considered in the fit of the D$^{0}$-meson $p_{\rm T}$ distribution in 20--40\% Pb--Pb collisions. No significant difference is observed in the resulting $v_{2}^{e \leftarrow c}$ with respect to the one considering only the statistical uncertainty in the fit.  For further investigation, The BAMPS result on the \pT distribution of D mesons  at $|y| <$ 0.8 in 20--40\% Pb--Pb collisions at \snn = 2.76 TeV~\cite{Uphoff:2014hza,Uphoff:2012gb} is also used to compute the $v_{2}^{e \leftarrow c}$. The relative difference of the obtained $v_{2}^{e \leftarrow c}$ using the estimated \pT distribution of D$^{0}$ mesons and the BAMPS result, which increases from 1\% to 20\% in the  interval  2 $< p_{\rm T} <$ 8 GeV/$c$, is included in the systematic uncertainty. 

The effect of the D$^{0}$-meson \vtwo estimation in 20--40\% Pb--Pb collisions using the arithmetic average of the D$^{0}$-meson \vtwo measurements in the 10--30\% and 30--50\% centrality classes is investigated in this analysis.
For this purpose, the trend of the unidentified charged particle \vtwo as a function of the average number of binary collisions ($\langle N_{coll }\rangle$)~\cite{centralitypaper} is assumed to be the same as the one for D$^{0}$ mesons. 
The  \vtwo as a function of  $\langle N_{coll }\rangle$ is obtained from a parametrisation of the centrality-dependent \vtwo measurement of unidentified charged particles  integrated over the interval 0.2 $< p_{\rm T} <$ 5 GeV/$c$~\cite{lighthadronv23}. The corresponding result exhibits a linear dependence between \vtwo and $\langle N_{coll }\rangle$ with a negative slope for  $\langle N_{coll }\rangle >$ 220. For comparison, the parametrisation is also obtained from the centrality-dependent \vtwo measurement of unidentified charged particles  integrated over the interval 10 $< p_{\rm T} <$ 20 GeV/$c$~\cite{ALICEChpartv2}. The  linear dependence between \vtwo and $\langle N_{coll }\rangle$ is the same as the one obtained for particles in a lower \pT interval.
The D$^{0}$-meson \vtwo is then obtained by the average of the D$^{0}$-meson \vtwo in the 10--30\% and 30--50\% centrality classes weighted by the \vtwo coefficients of the corresponding $\langle N_{coll }\rangle$ values~\cite{centralitypaper}. 
The relative difference of the obtained D$^{0}$-meson \vtwo with respect to the one obtained with the arithmetic average is negligible for $p_{\rm T} <$ 8 GeV/$c$ and its average is 19\% for $p_{\rm T} >$ 8 GeV/$c$, which is still compatible within uncertainties.
The $v_{2}^{e \leftarrow c}$ coefficients obtained with the two approaches show a relative difference of 2\% in the range 2 $< p_{\rm T} <$ 8 GeV/$c$. 
This deviation is considered as a consequence of statistical fluctuations in the D$^{0}$-meson \vtwo measurement for $p_{\rm T} >$ 8 GeV/$c$  and thus no systematic uncertainty is assigned for this effect.

In order to investigate the impact of the assumption of the particle mass ordering of the elliptic flow~\cite{Evidencehydrobis} used to determine the systematic uncertainty from the B feed-down subtraction in the D$^{0}$-meson \vtwo measurement, one can assume that the  \vtwo of prompt D mesons from B-meson decays should be between zero and the unidentified charged particle $v_{2}$.  The unidentified charged particle \vtwo in 20--40\% Pb--Pb collisions is obtained by the average of the \vtwo measurements in the 20--30\% and 30--40\% centrality classes~\cite{ALICEChpartv2} weighted by the corresponding $\langle N_{coll }\rangle$ values. 
The \vtwo coefficients of  prompt D$^{0}$ mesons and unidentified charged particles are compatible within uncertainties as well as the $v_{2}^{e \leftarrow c}$ obtained with these two results. Therefore, the lower limit of the systematic uncertainty from the B feed-down subtraction can be positioned at the central values of the prompt D-meson \vtwo and $v_{2}^{e \leftarrow c}$ without strictly considering that the B-meson \vtwo is expected to be lower than the D-meson $v_{2}$.

As a consequence of the \pT interval (2 $<  p_{\rm T} <$ 16 GeV/$c$) of the D$^{0}$-meson \vtwo and $p_{\rm T}$-differential yield measurements, the  $v_{2}^{e \leftarrow c}$ is obtained in the range 2 $< p_{\rm T} <$ 8 GeV/$c$.
The fraction of electrons with $p_{\rm T} >$ 2 GeV/$c$ that come from D$^{0}$ mesons with $p_{\rm T} <$ 2 GeV/$c$ is negligible according to PYTHIA simulations.
The effect of the \pT upper limit  of the D$^{0}$-meson measurements is studied by evaluating the $v_{2}^{e \leftarrow c}$ with extrapolation of the \pT and \vtwo  distributions of D$^{0}$ mesons up to 26 GeV/$c$.  The transverse momentum extrapolation is obtained from the power-law fit function shown in the left panel of Fig.~\ref{fig:CockInput}, while the impact of the \vtwo of D$^{0}$ mesons is estimated by explicitly setting its value, in the interval 16 $< p_{\rm T} <$ 26 GeV/$c$, to either zero, or constant at high $p_{\rm T}$, or maximum value of the prompt D$^{0}$-meson $v_{\rm 2}$ (shown in Fig.~\ref{fig:CockInput}). The highest relative difference in these three scenarios, which increases from 0.3\% to 40\% in the interval 2 $< p_{\rm T} <$ 8 GeV/$c$, is assigned as a conservative systematic uncertainty.

The effect of the mid-rapidity range of D$^{0}$ mesons is investigated by obtaining the $v_{2}^{e \leftarrow c}$ using the D$^{0}$-meson $p_{\rm T}$ distribution in the rapidity range $|y| <$ 1.6 as input for the simulation. The D$^{0}$-meson \vtwo is considered to be the same in this rapidity range, because no dependence on rapidity was observed in ALICE results on leptons from heavy-flavour hadron decays~\cite{Adam:2016ssk,HFERAA} as discussed previously. The relative difference of the obtained $v_{2}^{e \leftarrow c}$ with respect to the one using the D$^{0}$-meson $p_{\rm T}$ distribution in the rapidity range $|y| <$ 0.8 is negligible and thus no additional systematic uncertainty is considered due to the rapidity effect.

In this analysis, the \vtwo and shape of the $p_{\rm T}$-differential yields of charm hadrons are assumed to be the same as the ones measured for D$^{0}$ mesons. 
This is justified by the fact that the  \vtwo coefficients of  D$^{0}$, D$^{+}$ and D$^{*+}$ mesons are compatible within uncertainties in 30--50\% Pb--Pb collisions at \snn = 2.76 TeV~\cite{Abelev:2013lca}, also the prompt D$_ {s}^{+}$ \vtwo is compatible within uncertainties with the prompt non-strange D meson \vtwo in 30--50\%  Pb--Pb collisions at \snn = 5.02 TeV~\cite{Acharya:2017qps}. In addition,   the ratios of the yields of D$^{+}$/D$^{0}$ and D$^{*+}$/D$^{0}$  were observed to be constant within uncertainties in pp collisions at \s = 7 TeV and no modification of the ratios was observed within uncertainties in central and semi-central Pb--Pb collisions at \snn = 2.76 TeV~\cite{Adam:2015sza}. 
A possible hint for an enhancement of the D$_ {s}^{+}$/D$^{0}$ ratio is observed in 0--10\% Pb--Pb collisions at \snn = 2.76 TeV~\cite{Adam2016}, but the current uncertainties do not allow for a conclusion. 
The effect of  different decay kinematics of  charm particles is estimated by simulating the \vtwo of electrons from combined D, D$^{*}$, D$_ {s}$, and $\Lambda_{c}$ particle decays taking into account the fraction of charm quarks that hadronise into these particles~\cite{AMSLER20081} and using the same simulation input as used in the analysis ($p_{\rm T}$-differential yield and \vtwo of D$^{0}$ mesons). The obtained $v_{2}^{e \leftarrow c}$ is compatible with the one using D$^{0}$-meson decay and thus no systematic uncertainty is considered due to this effect. 
In order to exemplify the impact of a possible production enhancement of  D$_ {s}^{+}$ and $\Lambda_{c}$ particles in Pb--Pb collisions with respect to pp collisions, their fragmentation fractions are increased by a factor 2 and 5, respectively, in the simulation of the combined charm meson $v_{2}$.  The relative difference of the obtained $v_{2}^{e \leftarrow c}$ and the one using D$^{0}$-meson decay is negligible for $p_{\rm T} <$ 3 GeV/$c$ and its average is 5\% for  $p_{\rm T} >$ 3 GeV/$c$.

Finally, the D$^{0}$-meson $v_{\rm 2}$ systematic uncertainty from data is summed in quadrature with other sources of systematic uncertainty that affect significantly the $v_{2}^{e \leftarrow c}$ estimation, which are the \pT distribution of D$^{0}$ mesons  and the limited \pT interval of the D$^{0}$-meson measurements. They are considered as uncorrelated since the effect from the \pT distribution of D$^{0}$ mesons is obtained with the BAMPS result and the effect from the limited \pT interval of the D$^{0}$-meson measurements is obtained by extrapolations. The term ``from data'' is maintained later in this paper to distinguish all sources of systematic uncertainty from the systematic uncertainty related to the  B feed-down subtraction of the D$^{0}$-meson $v_{\rm 2}$ measurement, which is shown separately.

\subsection{Elliptic flow of electrons from beauty-hadron decays}
\label{Sec:ElectronFromBeautyV2}

The \vtwo of electrons from beauty-hadron decays ($v_{2}^{e \leftarrow b}$) is  obtained from Eq.~\ref{eq:ElecFrombeautyV2} using the  $R$,  $v_{2}^{e \leftarrow c+b}$ and $v_{2}^{e \leftarrow c}$ results presented in their respective sections.

The three results are considered as statistically independent. First, the factor $R$ was measured in a different collision system (pp collisions) or obtained with calculations. Second, the $v_{2}^{e \leftarrow c}$ is obtained with a simulation using measurements of D$^{0}$ mesons reconstructed via the hadronic decay channel $\rm{D}^{0} \rightarrow \rm{K}^{-} \pi^{+}$ in a different centrality class than in the $v_{2}^{e \leftarrow c+b}$  measurement. 

Even though the systematic uncertainties of the $R$,  $v_{2}^{e \leftarrow c+b}$ and $v_{2}^{e \leftarrow c}$ results might be partially correlated, especially concerning the particle identification selection criteria, the limited public information prevents  a more accurate treatment of these uncertainties. Therefore, they are assumed to be uncorrelated as a conservative estimation. As an example of the effect of a possible overestimation, if the systematic uncertainties of the $v_{2}^{e \leftarrow c}$ and $v_{2}^{e \leftarrow c+b}$ results decrease by 30\% in the interval 2 $< p_{\rm T} <$ 8 GeV/$c$, the systematic uncertainty  from data of the $v_{2}^{e \leftarrow b}$ result is expected to decrease by approximately 24\%.

Therefore, the statistical and systematic uncertainties of the $R$,  $v_{2}^{e \leftarrow c+b}$ and $v_{2}^{e \leftarrow c}$ results are propagated as independent variables. 
The $v_{2}^{e \leftarrow b}$ systematic uncertainties from data and from the B feed-down subtraction are asymmetric as a consequence of the systematic uncertainty asymmetry of the measurements used in this analysis. 
The systematic uncertainty from data is evaluated according to the method described in~\cite{DAgostini:2004yu}, where the positive and negative deviations are obtained separately and their average is added in quadrature. For verification, the alternative approach presented in~\cite{Barlow:2003sg} is also applied in this analysis. No significant difference between these methods is observed. Since the asymmetry of the  systematic  uncertainty from the B feed-down subtraction only comes  from the $v_{2}^{e \leftarrow c}$ result, the limits of the $v_{2}^{e \leftarrow b}$ systematic uncertainty are the deviations  resulting from the upper and lower limits of the $v_{2}^{e \leftarrow c}$ systematic uncertainty.

\section{Results}
\label{Sec:Results}

\begin{figure}
\centering
  \includegraphics[height=0.355\textwidth]{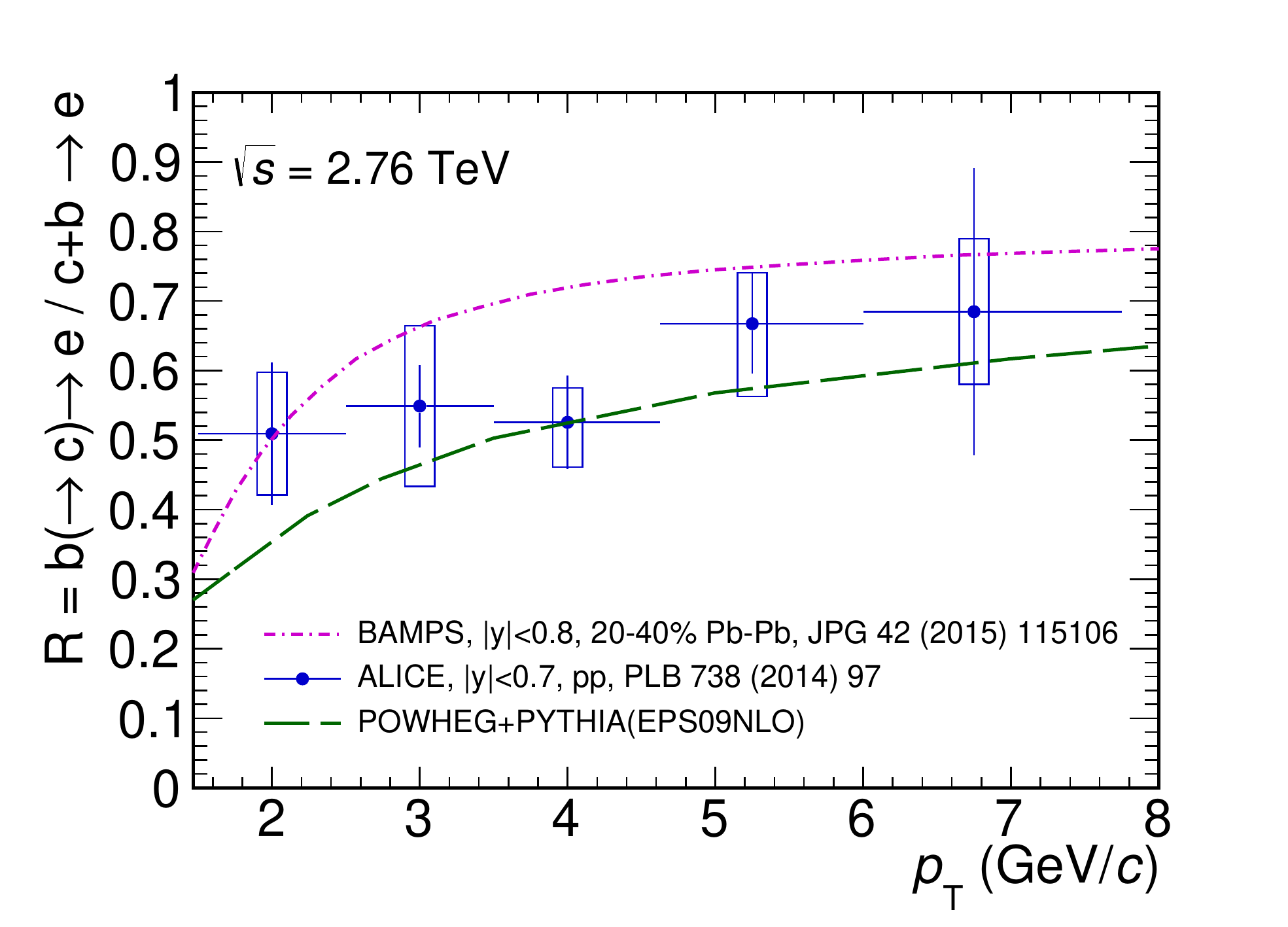}
    \includegraphics[height=0.355\textwidth]{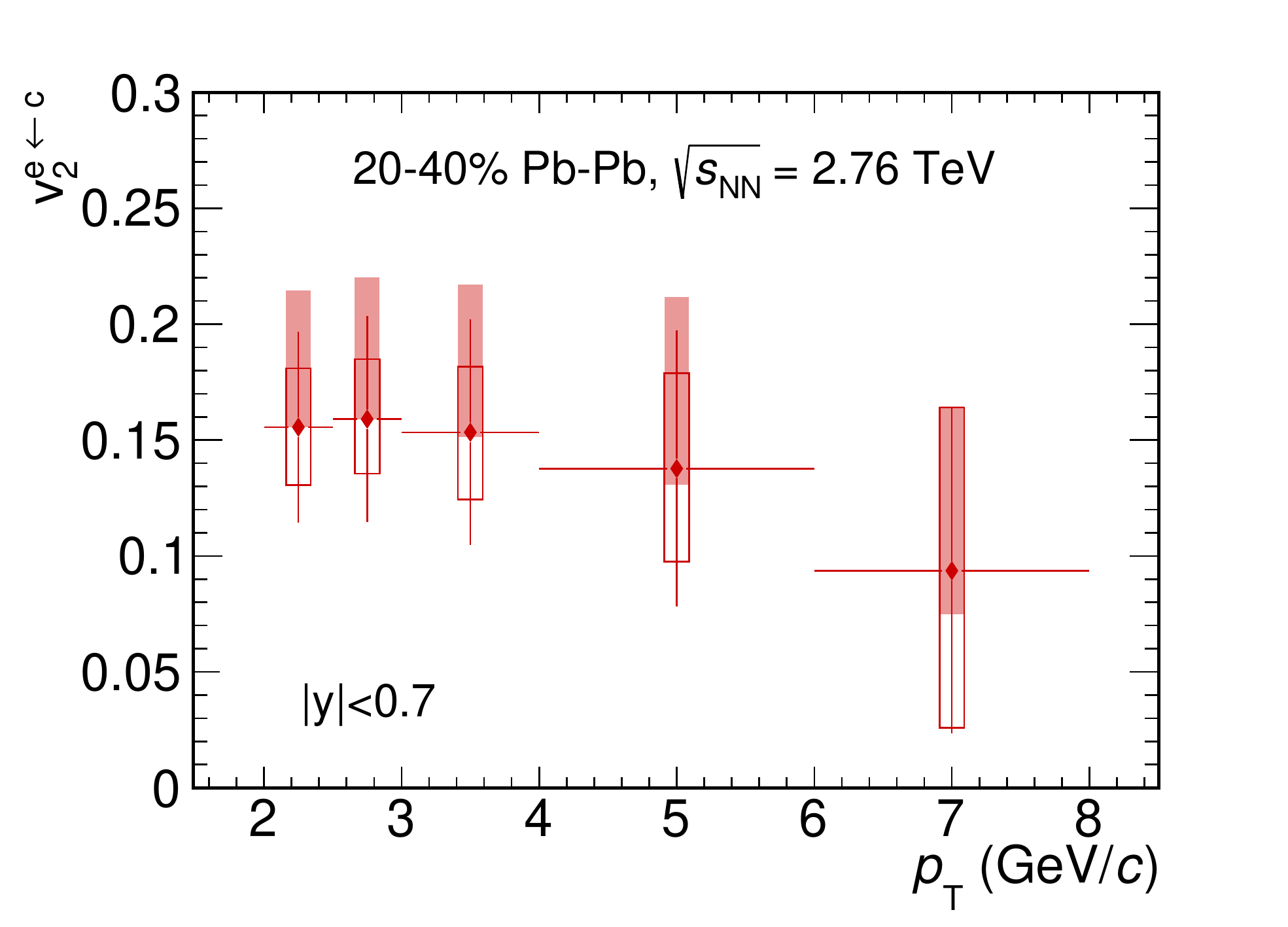}
  \caption{Left:  Relative contribution of electrons from beauty-hadron decays to the yield of electrons from heavy-flavour hadron decays at \s = 2.76 TeV obtained with POWHEG+PYTHIA at NLO accuracy  using  EPS09 NLO nPDFs. The result is compared with the $R$ in pp collisions at  \s = 2.76 TeV measured by the ALICE collaboration using the electron-hadron azimuthal correlation technique~\cite{Abelev:2014hla,2016507} and with the BAMPS result in 20--40\%  Pb--Pb collisions at \snn = 2.76 TeV~\cite{Uphoff:2014hza}. The statistical and systematic uncertainties of the $R$ coefficient obtained with POWHEG+PYTHIA and from the BAMPS  model are zero.  Right: Elliptic flow of electrons from charm-hadron decays at mid-rapidity  in 20--40\%  Pb--Pb collisions at \snn = 2.76 TeV estimated using a MC simulation with PYTHIA based on ALICE results~\cite{ALICEDRPbPb,Abelev:2014ipa}. 
The vertical error bars represent the statistical uncertainties and the horizontal error bars indicate the bin widths. The empty and filled boxes represent the systematic uncertainties from data and from the B feed-down subtraction, respectively, in the D$^0$-meson \vtwo measurement~\cite{Abelev:2014ipa}.}
   \label{fig:R_cbEv2_cEv2}
\end{figure}

The relative contribution of electrons from beauty-hadron decays to the yield of electrons from heavy-flavour hadron decays at \s = 2.76 TeV obtained with POWHEG+  PYTHIA at NLO accuracy using EPS09 NLO nPDFs is shown in the left panel of Fig.~\ref{fig:R_cbEv2_cEv2}. The result is compared with the $R$ in pp collisions at  \s = 2.76 TeV measured by the ALICE collaboration using the electron-hadron azimuthal correlation technique~\cite{Abelev:2014hla,2016507} and with the BAMPS result in 20--40\%  Pb--Pb collisions at \snn = 2.76 TeV~\cite{Uphoff:2014hza}. The comparison shows that $R$ is higher when in-medium effects are present, which is consistent with the expectation of the mass hierarchy of the energy loss of charm and beauty quarks in the medium.
The right panel of Fig.~\ref{fig:R_cbEv2_cEv2} shows the \vtwo of electrons from charm-hadron decays at mid-rapidity in 20--40\%  Pb--Pb collisions at \snn = 2.76 TeV obtained with a MC simulation with PYTHIA using as input the $p_{\rm T}$-differential yield and \vtwo distributions of D$^{0}$ mesons  in the same collision system. A positive \vtwo of electrons from charm-hadron decays is found in all \pT intervals, with a  maximum significance of 3.2$\sigma_{-}$, where $\sigma_{-}$ is the combined statistical and systematic uncertainties of the lower limit,  in the interval 2 $<  p_{\rm T} <$ 3 GeV/$c$.

\begin{figure*}
\centering
  \includegraphics[height=0.45\textwidth]{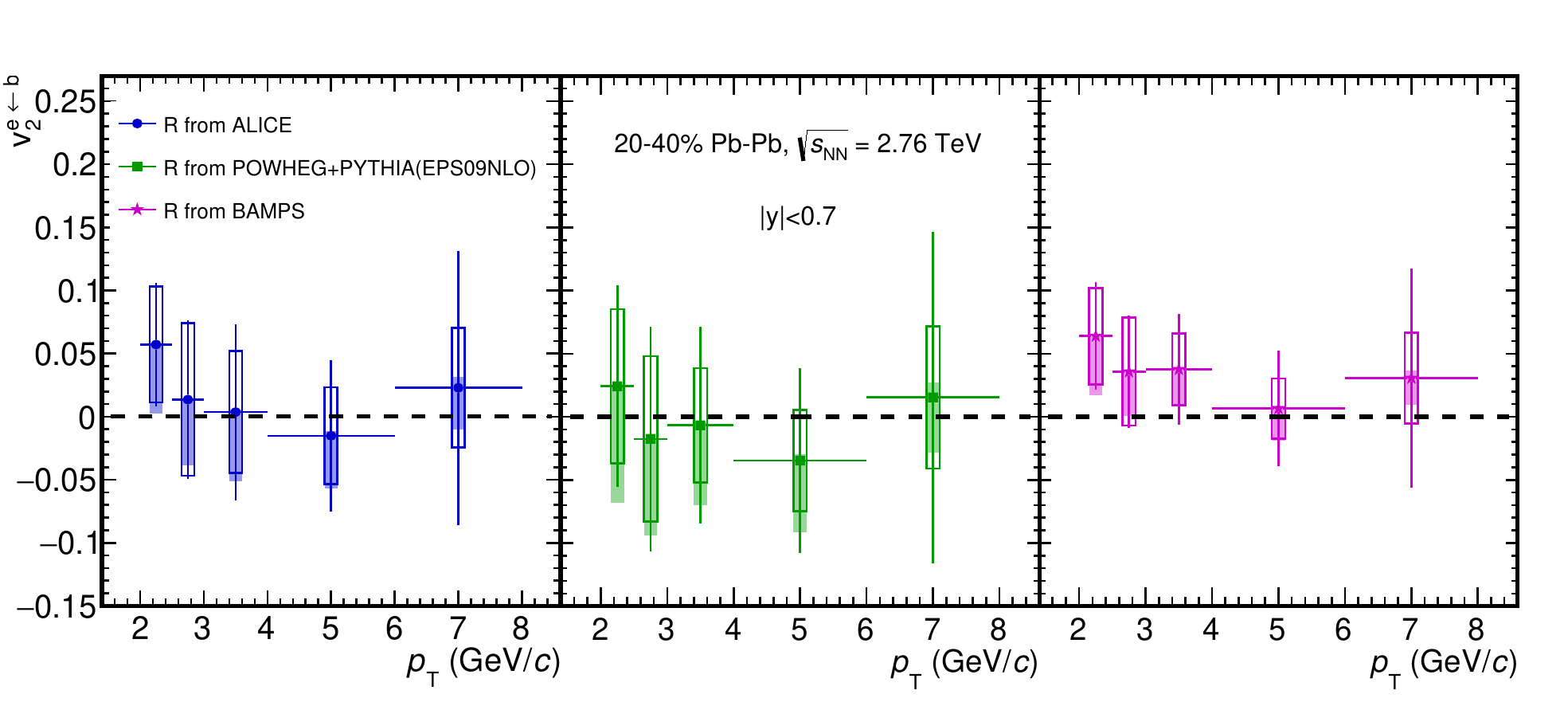}
  \caption{Elliptic flow of electrons from beauty-hadron decays at mid-rapidity in 20--40\%  Pb--Pb collisions at \snn = 2.76 TeV using the  approaches of the factor $R$~\cite{Abelev:2014hla,2016507,Uphoff:2014hza} shown in the left panel of Fig.~\ref{fig:R_cbEv2_cEv2}. The vertical error bars represent the statistical uncertainties and the horizontal error bars indicate the bin widths.  The empty and filled boxes represent the systematic uncertainties from data and from the B feed-down subtraction, respectively, in the D$^0$-meson \vtwo measurement~\cite{Abelev:2014ipa}.}
   \label{fig:ElecFromBeautyv2_comparisonR_20161014.pdf}
\end{figure*}

The \vtwo coefficients of electrons from beauty-hadron decays in 20--40\%  Pb--Pb collisions at \snn = 2.76 TeV obtained with different approaches of the factor $R$ (left panel of Fig.~\ref{fig:R_cbEv2_cEv2}) are shown in Fig.~\ref{fig:ElecFromBeautyv2_comparisonR_20161014.pdf}.
The result computed with  the coefficient $R$ in pp collisions is an estimation of the minimum value, as discussed previously. The \vtwo of electrons from beauty-hadron decays in 20--40\% Pb--Pb collisions at \snn = 2.76 TeV is compatible with zero within approximately 1$\sigma$  of the total uncertainty, obtained by summing in quadrature the different uncertainty contributions, in all \pT intervals and different  $R$ coefficients. However, the large statistical and systematic uncertainties prevent  a definite conclusion. 
The result is consistent with the measured \vtwo of non-prompt J/$\psi$ mesons from $\mathrm{B}$-hadron decays in 10--60\% Pb--Pb collisions at \snn = 2.76 TeV~\cite{Khachatryan:2016ypw}, which is also compatible with zero within uncertainties.

\begin{figure}
\centering
    \includegraphics[height=0.45\textwidth]{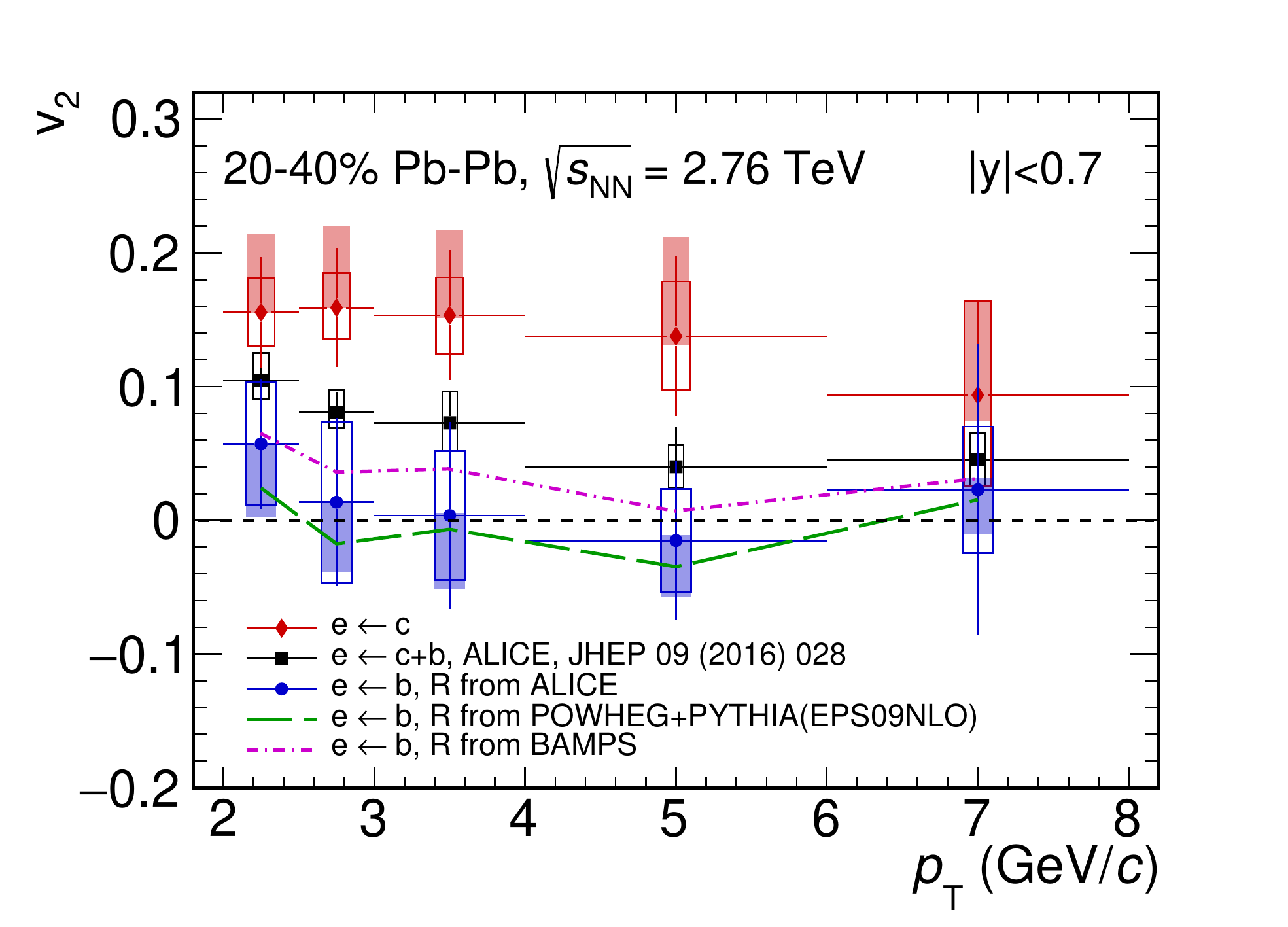}
  \caption{Elliptic flow of electrons from charm- and beauty-hadron decays, inclusive~\cite{Adam:2016ssk} and separated, in 20--40\%  Pb--Pb collisions at \snn = 2.76 TeV. The vertical error bars represent the statistical uncertainties and the horizontal error bars indicate the bin widths.  The empty and filled boxes represent the systematic uncertainties from data and from the B feed-down subtraction, respectively, in the D$^0$-meson \vtwo measurement~\cite{Abelev:2014ipa}.}  
   \label{fig:v2_ComparisonOfAllDecays}
\end{figure}

Figure~\ref{fig:v2_ComparisonOfAllDecays} shows the \vtwo of electrons from charm- and beauty-hadron decays, inclusive~\cite{Adam:2016ssk} and separated, in 20--40\%  Pb--Pb collisions at \snn = 2.76 TeV.
The \vtwo of electrons from beauty-hadron decays is lower than the \vtwo of electrons from charm-hadron decays, although they are compatible within uncertainties. 
The average of the \vtwo coefficients of electrons from  charm- and beauty-hadron decays obtained in the interval 2 $<  p_{\rm T} <$ 8 GeV/$c$ are listed in Table~\ref{tab:IntV2}. Because of the asymmetric uncertainties, the average is obtained numerically with an iterative sum of the likelihood functions parametrised by variable-width Gaussians~\cite{Barlow:2004wg,Barlow:2003sg}.
The standard deviation, which is the combination of  statistical and systematic uncertainties, is assumed to vary linearly. 
The maximum value of the  summed likelihood function corresponds to the average $v_{2}$, while the points at which the function is -0.5 correspond to the  lower and upper limits of the total uncertainty.

\begin{table}
\renewcommand{\arraystretch}{1.5}
\setlength{\tabcolsep}{4pt}
\centering
\begin{tabular}{ c l c }
\hline  \hline
Result & R approach & Average \vtwo \\
\hline \hline
$e \leftarrow c$ &  -- & $\rm 0.150_{\rm - 0.028}^{\rm + 0.034}$ \\
$e \leftarrow b$ & ALICE & $\rm 0.014 _{\rm - 0.042}^{\rm + 0.039}$  \\
$e \leftarrow b$ & POWHEG+PYTHIA(EPS09NLO) & $\rm -0.010 _{\rm - 0.052}^{\rm + 0.047}$  \\
$e \leftarrow b$ & BAMPS & $\rm 0.032_{\rm - 0.030}^{\rm + 0.028}$ \\
\hline  \hline
\end{tabular}
  \caption{Average of the \vtwo coefficients of electrons from  charm- and beauty-hadron decays obtained in the  transverse momentum interval  2 $<  p_{\rm T} <$ 8 GeV/$c$ in 20--40\%  Pb--Pb collisions at \snn = 2.76 TeV. The reported errors are the combined   statistical and systematic uncertainties. See text for more details.}
\label{tab:IntV2}
\end{table}

\section{Conclusions}
\label{Sec:Conclusions}
 
We presented a method to subtract the charm quark contribution in the elliptic flow of electrons from heavy-flavour hadron decays.
The \vtwo of electrons from charm-hadron decays was estimated using a MC simulation of D$^0$-meson  decays  into electrons with PYTHIA, based on  measurements of the $p_{\rm T}$-differential yield  and \vtwo of D$^{0}$ mesons  in Pb--Pb collisions at \snn = 2.76 TeV by  ALICE.
A positive \vtwo of electrons from charm-hadron decays is found with a maximum significance of 3.2$\sigma_{-}$ in the interval 2 $<  p_{\rm T} <$ 3 GeV/$c$.
The computed \vtwo  of electrons from charm-hadron decays was  then subtracted from the  \vtwo of electrons from heavy-flavour hadron decays in  20--40\%  Pb--Pb collisions at \snn = 2.76 TeV measured by the ALICE collaboration. 
The subtraction was weighted by the relative contribution of electrons from beauty-hadron decays to the yield of electrons from heavy-flavour hadron decays. Since this observable is not measured in Pb--Pb collisions, three different approaches were used as estimations in the analysis. 
The resulting \vtwo of electrons from beauty-hadron decays in 20--40\%  Pb--Pb collisions at \snn = 2.76 TeV from the subtraction is compatible with zero within approximately 1$\sigma$  of the total uncertainty in all \pT intervals and different  approaches of the relative contribution of electrons from beauty-hadron decays to the yield of electrons from heavy-flavour hadron decays. However, the large statistical and systematic uncertainties prevent a definite conclusion. The \vtwo of electrons from beauty-hadron decays is found to be lower than the \vtwo of electrons from charm-hadron decays.

 \section{Outlook}
 \label{Sec:Outlook}

In the presented method, the elliptic flow of electrons from beauty-hadron decays can be determined by using three observables that have largely been measured at the LHC and RHIC.
Based on  available results of these observables, the procedure was applied using measurements  performed by the ALICE collaboration. The method  demonstrated to be effective; however, the current statistical and systematic uncertainties of the ALICE results prevent a definite conclusion whether the collective motion of the medium constituents influences beauty quarks. 
A better accuracy of the results on heavy-flavour particles has been achieved in measurements in Pb--Pb collisions at \snn = 5.02 TeV~\cite{Acharya:2017qps} and it is expected to be further improved with the ALICE upgrade, which is foreseen to start in 2019. 

In particular, the upgrade of the Inner Tracking System (ITS) detector will improve the determination of the distance of closest approach to the primary vertex, momentum resolution and readout rate capabilities~\cite{0954-3899-41-8-087002}. These improvements will allow for more precise measurements of D mesons down to low transverse momenta and for reducing the systematic uncertainties  from data and from the B feed-down subtraction. 
The latter will be possible with the direct measurement of the fraction of prompt D mesons and D mesons from B-meson decays, which is expected to be accessible with relative statistical and systematic uncertainties smaller than 1\% and 5\%~\cite{0954-3899-41-8-087002}, respectively, for prompt D$^{0}$ mesons.
In addition, the ITS upgrade will enable the tracking of electrons down to approximately 0.05 GeV/$c$ and enhance the capability to separate prompt from displaced electrons~\cite{0954-3899-41-8-087002}, improving the reconstruction of electrons that do not originate from heavy-flavour hadron decays needed for the background subtraction. Moreover, the systematic uncertainty of the elliptic flow of electrons from beauty-hadron decays can be further improved by taking into account correlations among different contributions.

The capability of the heavy-flavour measurements will also enhance with the increase of luminosity. For instance, the current relative statistical uncertainty of the D-meson \vtwo measurement in Pb--Pb collisions is 10\% for an integrated luminosity of 0.1 nb$^{-1}$, while it is expected to be 0.2\% for a scenario with an integrated luminosity of 10 nb$^{-1}$~\cite{0954-3899-41-8-087002}. Also the elliptic flow coefficients of D$_{s}$ and $\Lambda_{c}$  particles are expected to be achievable with a relative statistical uncertainty of 8\% and 20\%~\cite{0954-3899-41-8-087002}, respectively. 

Therefore, the presented method can be used to extract the elliptic flow of electrons from beauty-hadron decays with better precision with future measurements of the three needed observables.

\section*{Acknowledgements}
We would like to thank Carsten Greiner and Florian Senzel for providing the BAMPS results, as well as Francesco Prino for fruitful discussions.
We are grateful for the support of the Deutsche Forschungsgemeinschaft (DFG)    through the Research Training Group ``GRK 2149: Strong and Weak Interactions - from Hadrons to Dark Matter''; Bundesministerium f\"ur Bildung und Forschung (BMBF) under the project number  05P15PMCA1; Conselho Nacional de Desenvolvimento Cient\'ifico e Tecnol\'ogico (CNPq); and Funda\c{c}\~ao de Amparo \`a Pesquisa do Estado de S\~ao Paulo (FAPESP).


\bibliography{biblio}
\bibliographystyle{spmpsci2}       

\end{document}